# Angular Integral Autocorrelation for Speed Estimation in Shear Wave Elastography

Hamidreza Asemani[1,2], Irteza Enan Kabir[1], Juvenal Ormachea[3], Marvin M Doyley[2], Jannick P Rolland[1], Kevin J Parker[2]

[1] Institute of Optics, University of Rochester, 480 Intercampus Drive, Box 270186, Rochester, NY 14627, United States of America
[2] Department of Electrical and Computer Engineering, University of Rochester, 724 Computer Studies Building, Box 270231, Rochester, NY 14627, United States of America
[3] Verasonics, Inc., 11335 NE122nd Way, Suite 100 98034 Kirkland, WA, United States of America
E-mail: kevin.parker@rochester.edu

**Abstract**
The utilization of a reverberant shear wave field in shear wave elastography has emerged as a promising technique for achieving robust shear wave speed (SWS) estimation. However, accurately measuring SWS within such a complex wave field remains challenging. This study introduces an advanced autocorrelation estimator based on angular integration known as the angular integral autocorrelation (AIA) approach to address this issue. The AIA approach incorporates all the autocorrelation data from various angles during measurements, resulting in enhanced robustness to both noise and imperfect distributions in SWS estimation. The effectiveness of the AIA estimator for SWS estimation is first validated using a k-Wave simulation of a stiff branching tube in a uniform background. The results demonstrate that the AIA estimator, when compared with simple autocorrelation approaches, improves both the accuracy of the estimated SWS ratio between the stiff branching structure and background and the signal-to-noise ratio (SNR) in estimating SWS. To further evaluate the effectiveness of the AIA estimator, ultrasound elastography experiments, MRI experiments, and OCT studies across a range of different excitation frequencies are conducted on a series of tissues and phantoms, including *in vivo* scans. The results verify the capacity of the AIA approach to enhance the accuracy of SWS estimation and SNR, even within an imperfect reverberant shear wave field. Compared to simple autocorrelation approaches, the AIA approach can also successfully visualize and define lesions while significantly improving the estimated SWS and SNR in homogeneous background materials, thus providing improved elastic contrast between structures within the scans. These findings demonstrate the robustness and effectiveness of the AIA approach across a wide range of applications, including ultrasound elastography, MRE, and OCE, for accurately identifying the elastic properties of biological tissues in diverse excitation scenarios.

## 1. Introduction

Shear wave elastography (SWE) has been increasingly utilized in medical imaging for its ability to visualize tissue stiffness *in vivo*, combined with its capability for diagnostic disease assessment (Davis *et al* 2019, Leartprapun and Adie 2023, Venkates *et al* 2013). By measuring the speed of the shear waves, SWE provides valuable information about tissue elasticity (Zvietcovich *et al* 2016, McGarry *et al* 2022) and viscoelasticity (Babaei *et al* 2021, Parker and Ormachea 2021, Poul *et al* 2022). This technology has found applications in the diagnosis of different diseases such as liver fibrosis (Ferraioli *et al* 2014, Poul and Parker 2021), breast cancer

(Chang *et al* 2013), skin lesions (Es'haghian *et al* 2015), and brain diseases such as Alzheimer's (Ge *et al* 2022a).

SWE can be enabled in different modalities, including ultrasound (Doyley and Parker 2014, Ormachea *et al* 2018), magnetic resonance imaging (MRI) (Fovargue *et al* 2018, Herthum *et al* 2022, Sack 2023), and optical coherence tomography (OCT) (Wang and Larin 2015, Zvietcovich *et al* 2017). These modalities offer complementary advantages in terms of imaging depth, spatial resolution, field of view, and tissue characterization, expanding the capabilities of SWE for clinical applications and research studies.

Ultrasound elastography offers several advantages including almost real-time imaging, low imaging costs, a wide field of view, and significant penetration depth, making ultrasound elastography well-suited for a wide range of clinical applications with good spatial resolution (Ormachea and Parker 2020, Sigrist *et al* 2017). Magnetic resonance elastography (MRE), with its excellent soft tissue contrast and ability to assess large tissue volumes, provides three-directional and three-dimensional elastograms and is well suited for tissue characterization in various clinical applications, particularly in liver fibrosis assessment and brain studies (Low *et al* 2016, Mariappan *et al* 2010). Optical coherence elastography (OCE) on the other hand provides high-resolution 3D imaging capabilities and is ideal for applications in ophthalmology and dermatology (Kennedy *et al* 2014, Zvietcovich and Larin 2022).

There are numerous estimators available in elastography, the broad categories of these are outlined in a review paper by Doyley (2012). In shear wave elastography, there are several approaches including the phase gradient method (Ormachea and Parker 2021), the time-of-flight method (Zvietcovich *et al* 2019a), the Fourier estimator (Zvietcovich *et al* 2017), the viscoelastic wave inversion method (Barnhill *et al* 2018), wave attenuation estimator (Ramier *et al* 2019), and the autocorrelation estimator (Parker *et al* 2017, Zvietcovich *et al* 2019b). Autocorrelation estimators are widely recognized as robust and effective approaches for SWS measurement in various elastography scans, especially for reverberant shear wave fields (Parker *et al* 2017). However, choosing the appropriate autocorrelation estimator and autocorrelation parameters for a particular elastography application requires careful consideration of various factors, such as the imaging system, tissue properties, experimental conditions, and mechanical excitation (Zvietcovich *et al* 2019b).

Despite the developments in SWE, accurately measuring SWS in elastography scans is still challenging. For example, generating a fully reverberant shear wave field in 3D typically requires several excitation sources plus reflections, this is not always achieved in real experiments. Imperfect reverberant shear wave fields that exhibit a dominant direction or source, especially in proximity to an external source can be found. To address this issue, we introduce a novel autocorrelation estimator based on angular integration for SWS measurement within the general framework established for reverberant shear wave fields. The effectiveness of the angular integral autocorrelation (AIA) estimator is examined using numerical simulations and experimental data from ultrasound elastography, MRE, and OCE imaging systems.

This paper is organized to first review the fundamental equations for reverberant shear wave fields and earlier and simple estimators of shear wavelength in these fields. Next, the AIA estimator is introduced. These estimators are then tested in numerical simulations of shear wave fields and in experiments using ultrasound, MRI, and OCT imaging platforms across a variety of conditions and elastic biomaterials. Final comparisons illustrate the superiority of the proposed AIA estimator showing its robust performance even in the presence of non-ideal conditions.

## 2. Theory

When shear waves are generated by multiple excitation sources or wave reflections happen at tissue boundaries, shear wave interferences generate a reverberant shear wave field or diffuse shear wave field (Pierce, 1981, Parker and Maye 1984, Parker *et al* 2017). A fully reverberant shear wave field can be characterized as the superposition of several planar shear waves propagating in random directions with the same wavenumber $k$ and angular frequency $\omega_0$. However if the number of plane waves is inadequate or the wave traveling directions are non-random, imperfect, or non-fully reverberant, shear wave fields are produced.

The particle velocity of a fully reverberant shear wave field $\boldsymbol{V}(\boldsymbol{\varepsilon}, t)$ can be defined as:

$$\boldsymbol{V}(\boldsymbol{\varepsilon}, t) = \sum_{q,l} \hat{\boldsymbol{n}}_{ql} \, v_{ql} \, e^{i(k\hat{\boldsymbol{n}}_q \cdot \boldsymbol{\varepsilon} - \omega_0 t)}, \tag{1}$$

where $\boldsymbol{\varepsilon}$ represents the position vector, $t$ denotes time, and the indices $q$ and $l$ refer to realizations of the random unit vectors $\hat{\boldsymbol{n}}_q$ and $\hat{\boldsymbol{n}}_{ql}$, respectively. The vector $\hat{\boldsymbol{n}}_q$ represents the random direction of wave propagation, while $\hat{\boldsymbol{n}}_{ql}$ represents a random unit vector indicating the direction of particle motion. $v_{ql}$ is an independent, identically distributed random variable representing the magnitude of the particle velocity within a realization.

It is worth mentioning that in transversal shear wave fields, the wave propagation direction is perpendicular to the particle motion which implies that $\hat{\boldsymbol{n}}_{ql} \cdot \hat{\boldsymbol{n}}_q = 0$.

Typically, elastography modalities such as ultrasound elastography, MRE, and OCE are able to measure the particle velocity along a sensor axis (note that MRE can determine displacements along three axes at the expense of longer scan times). It is common practice to designate the sensor axis as the *z*-axis. Consequently, for the scalar velocity field in the *z*-direction $V_z(\boldsymbol{\varepsilon}, t)$ we have:

$$V_z(\boldsymbol{\varepsilon}, t) = \boldsymbol{V}(\boldsymbol{\varepsilon}, t) \cdot \hat{\boldsymbol{e}}_z = \sum_{q,l} n_{ql_z} \, v_{ql} \, e^{i(k\hat{\boldsymbol{n}}_q \cdot \boldsymbol{\varepsilon} - \omega_0 t)}, \tag{2}$$

where $\hat{\boldsymbol{e}}_z$ is the *z*-direction unit vector and $n_{ql_z} = \hat{\boldsymbol{n}}_{ql} \cdot \hat{\boldsymbol{e}}_z$ is a scalar random variable. As discussed in the introduction section, for elastography purposes the SWS or the wave number of the reverberant shear wave field should be calculated. To summarize equation (2) and calculate the $k$ number, one approach derives an autocorrelation estimation. So, by considering the autocorrelation function of equation (2) both in space and time, $B_{V_z V_z}(\Delta \boldsymbol{\varepsilon}, \Delta t)$, we derive the following expression:

$$B_{V_z V_z}(\Delta \boldsymbol{\varepsilon}, \Delta t) = E\{V_z(\boldsymbol{\varepsilon}, t) V_z^*(\boldsymbol{\varepsilon} + \Delta \boldsymbol{\varepsilon}, t + \Delta t)\}, \tag{3}$$

where $E$ represents an ensemble average, and the asterisk ($*$) denotes the complex conjugate. By substituting equation (2) into equation (3), we have:

$$B_{V_z V_z}(\Delta \boldsymbol{\varepsilon}, \Delta t) = E\left\{\left(\sum_{q,l} n_{ql_z} \, v_{ql} \, e^{i(k\hat{\boldsymbol{n}}_q \cdot \boldsymbol{\varepsilon} - \omega_0 t)}\right) \times \left(\sum_{q',l'} n_{q'l'_z} \, v_{q'l'} \, e^{-i(k\hat{\boldsymbol{n}}_{q'} \cdot (\boldsymbol{\varepsilon} + \Delta \boldsymbol{\varepsilon}) - \omega_0(t + \Delta t))}\right)\right\}. \tag{4}$$

Many of the terms in this equation arise from independent realizations and therefore the cross terms are equal to zero. Then, by utilizing the spherical coordinate system described by Aleman-Castañeda *et al* (2021), equation (4) simplifies to:

$$B_{V_zV_z}(\Delta\boldsymbol{\varepsilon}, \Delta t) = \overline{v^2}\, e^{i\omega_0 \Delta t}\left\{\frac{\sin^2\theta_s}{2}\left[j_0(k\Delta\varepsilon) - \frac{j_1(k\Delta\varepsilon)}{k\Delta\varepsilon}\right] + \cos^2\theta_s\, \frac{j_1(k\Delta\varepsilon)}{k\Delta\varepsilon}\right\}, \quad (5)$$

where $\overline{v^2}$ is the ensemble average velocity-squared, $\theta_s$ is the angle of $\Delta\boldsymbol{\varepsilon}$ with respect to the sensor axis (considered as the z-axis), $j_0$ and $j_1$ are the first kind spherical Bessel function of zero order and first order, respectively. It is important to note that the vector $\Delta\boldsymbol{\varepsilon}$ is a function of its magnitude $\Delta\varepsilon$ and its angle $\theta_s$. The autocorrelation function $B_{V_zV_z}(\Delta\boldsymbol{\varepsilon}, \Delta t)$ clearly depends on the direction of $\Delta\boldsymbol{\varepsilon}$. In simple baseline autocorrelation estimation, $\Delta\boldsymbol{\varepsilon}$ is assumed to align with one of the Cartesian axes; in other words, we have $\Delta\varepsilon_x$, $\Delta\varepsilon_y$, and $\Delta\varepsilon_z$. The angle $\theta_s$ for $\Delta\varepsilon_x$ and $\Delta\varepsilon_y$ is equal to $\pi/2$ and for $\Delta\varepsilon_z$ is equal to zero. Therefore, the simple baseline autocorrelation functions are defined as:

$$B_{V_zV_z}(\Delta\varepsilon_x, \Delta t) = \frac{1}{2}\overline{v^2}\, e^{i\omega_0\Delta t}\left[j_0(k\Delta\varepsilon_x) - \frac{j_1(k\Delta\varepsilon_x)}{k\Delta\varepsilon_x}\right] \quad (6.a)$$

$$B_{V_zV_z}(\Delta\varepsilon_y, \Delta t) = \frac{1}{2}\overline{v^2}\, e^{i\omega_0\Delta t}\left[j_0(k\Delta\varepsilon_y) - \frac{j_1(k\Delta\varepsilon_y)}{k\Delta\varepsilon_y}\right] \quad (6.b)$$

$$B_{V_zV_z}(\Delta\varepsilon_z, \Delta t) = \overline{v^2}\, e^{i\omega_0\Delta t}\, \frac{j_1(k\Delta\varepsilon_z)}{k\Delta\varepsilon_z}. \quad (6.c)$$

The simple baseline autocorrelation estimators are written in terms of spherical Bessel functions, and the wave number $k$ can be calculated using curve fitting. However, it is important to note that when utilizing simple baseline autocorrelation estimators, the autocorrelation data from different directions is ignored, and at minimum, only the autocorrelation function from one direction of lag is considered. This limitation makes the simple baseline autocorrelation estimators highly sensitive to noise and uncertainty, as information from other directions is disregarded. Furthermore, any non-ideal weighting of the shear wave distribution across all solid angles can cause bias in the estimate from only one or two lag directions of the sampled autocorrelation function.

To address the issue, we introduce a novel approach based on angular integration of the autocorrelation function. This approach allows us to incorporate the autocorrelation data from various directions while eliminating the dependence of the autocorrelation function on the angle. By performing the angular integral of the autocorrelation function within a two-dimensional plane over the range of 0 to $2\pi$, we obtain the following expression:

$$B_{AI}(\Delta\rho, \Delta t) = \frac{1}{2\pi}\int_0^{2\pi} B_{V_zV_z}(\Delta\rho, \theta_s, \Delta t)d\theta_s, \quad (7)$$

where $B_{AI}(\Delta\rho, \Delta t)$ represents the AIA function and $\Delta\rho$ is the one-dimensional shift in the autocorrelation argument after integration around $\theta_s$. It is important to note that the AIA will reduce to different analytic formulas across different autocorrelation planes. Considering the sensor is aligned with the z-axis, for xz or yz autocorrelation planes, the AIA function can be measured by substituting equation (5) in equation (7) and calculating the integral as follows:

$$B_{AI_{xz}}(\Delta\rho, \Delta t) = B_{AI_{yz}}(\Delta\rho, \Delta t) = \frac{1}{4}\overline{v^2}\, e^{i\omega_0\Delta t}\left[j_0(k\Delta\rho) + \frac{j_1(k\Delta\rho)}{k\Delta\rho}\right], \quad (8)$$

where $B_{AI_{xz}}(\Delta\rho, \Delta t)$ and $B_{AI_{yz}}(\Delta\rho, \Delta t)$ represent AIA functions in the xz and yz planes respectively. These functions account for the autocorrelation data obtained from different directions within each respective plane. However, for the *xy* autocorrelation plane, the angle $\theta_s$ is always equal to $\pi/2$ and so we have the following equation:

$$B_{AI_{xy}}(\Delta\rho, \Delta t) = \frac{1}{2}\overline{v^2}\, e^{i\omega_0 \Delta t}\left[j_0(k\Delta\rho) - \frac{j_1(k\Delta\rho)}{k\Delta\rho}\right], \qquad (9)$$

where $B_{AI_{xy}}(\Delta\rho, \Delta t)$ is the AIA function in the *xy* plane.

These two descriptions for AIA are independent of the angle $\theta_s$, allowing for straightforward estimation of the wavenumber through curve fitting or other related methods. Moreover, by incorporating all the autocorrelation data from various angles in the calculating autocorrelation function, this method has the potential to be less sensitive to noise and imperfect distributions, resulting in enhanced robustness in SWS measurements.

## 3. Methods

In this study, we introduce AIA, which calculates the angular integral of the two-dimensional autocorrelations across a designated plane. To assess the efficiency of the AIA approach, some simulations were conducted using the k-Wave simulation toolbox in MATLAB (The MathWorks, Inc. Natick, MA, USA, 2022b). Subsequently, our proposed method was employed to measure SWS in different modalities, including ultrasound elastography, MRE, and OCE.

In order to provide a comprehensive understanding of the AIA estimator, we detail the algorithm utilized for shear wave measurements. Figure 1 provides an overview of the algorithm employed to measure the SWS. The case study in this figure is ultrasound elastography of a breast phantom with a stiff lesion. As depicted in figure 1, the initial dataset comprises the displacement (or velocity) field obtained from ultrasound elastography, representing the cross-section of an object over time. The displacement field is structured as a 3D real matrix including the depth or axial dimension, lateral dimension, and time dimension. In the first step, an autocorrelation estimation can be applied to the displacement field for each time frame. However, to reduce computational cost and enhance robustness, it is preferable to measure the autocorrelation function in the frequency domain. Therefore, the fast Fourier transform (FFT) is performed in the time domain, followed by the selection of the maximum intensity in the frequency domain, corresponding to the excitation frequency. The resulting data in this step is a 2D complex matrix, comprising both magnitude and phase components, as presented in figure 1.

After applying various filtering techniques including finite impulse response (FIR) filtering and median filtering to experimental data to reduce unwanted noise in the results and enhance the quality and reliability of the data, a 2D autocorrelation window with a defined window size and window step (jump) moves through the 2D matrix. The autocorrelation of each window is computed, and the real part of the results is considered for further analysis. As illustrated in figure 1, each autocorrelation window generates a 2D real matrix, where the $\Delta\boldsymbol{\varepsilon}$ values within this matrix are in Cartesian coordinates but can be converted to polar $(r, \theta)$ values using $\varepsilon_x = r\cos\theta$ and $\varepsilon_y = r\sin\theta$. To calculate the angular integral, we sum over the autocorrelation matrix across constant values of $r$, resulting in a one-dimensional autocorrelation function of $\Delta\rho$. Full curve fitting or other related methods can be utilized to estimate the unknown wavenumber $k$ or the SWS for each measured curve. This approach allows to increase the accuracy and reliability of the estimates for the SWS, as it considers the autocorrelation data from all angles and reduces the impact of noise and uncertainty. Finally, the SWS of the entire 2D cross-section is determined by combining the estimated SWS from all autocorrelation windows across the entire cross-section.

This algorithm can also be extended for SWS estimation in a 3D medium by dividing the 3D medium into 2D planes. The same approach can be applied to each plane, treating them as individual cross-sections, and then the SWS can be combined to obtain a 3D representation of the SWS distribution, including possible anisotropy (Aleman-Castañeda et al 2021). This adaptation allows for the assessment of SWS variations in different planes of the 3D medium, providing a more comprehensive understanding of the mechanical properties throughout the entire volume. In this study, alongside the comparison to ground truth, the SNR within homogeneous regions is calculated using the following equation (Ormachea and Parker 2021):

$$SNR = 10\ log_{10}\left(\frac{Average\ of\ shear\ wave\ speed}{Standard\ deviation\ of\ shear\ wave\ speed}\right). \qquad (10)$$

This metric can be used across each of the examples and modalities presented in the results, and is computed using the largest region of interest (ROI) that can be selected within each data set within the nominally homogeneous background or interior. The SNR measurement for background regions (homogeneous regions) is chosen because it could be applied to all our examples from all modalities, providing a consistent measure across all examples. Ideally, it should achieve high levels, characterized by a uniform mean and low standard deviation across a homogeneous background in low noise.

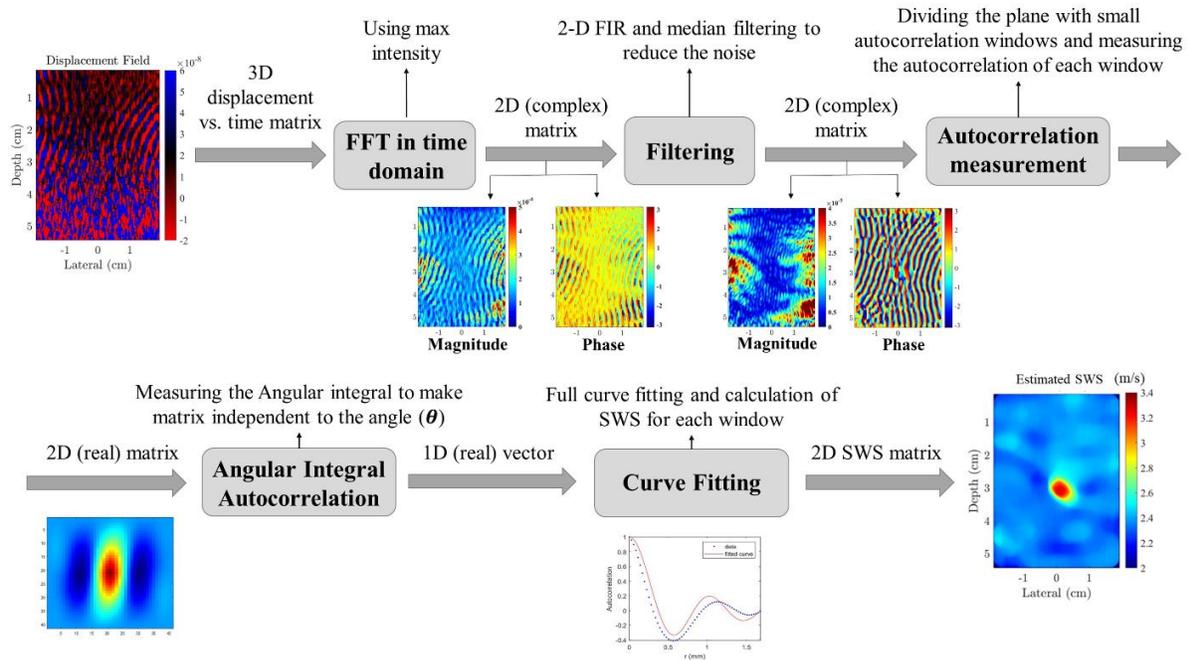

**Figure 1.** The algorithm for SWS estimation in ultrasound elastography of a phantom with a lesion.

### 3.1. Numerical simulations

To demonstrate the effectiveness of the AIA estimator for SWS estimation, a simulation was conducted in MATLAB (Treeby et al 2016). The simulation aimed to model real experimental conditions for different elastography modalities by considering various aspects of wave generation, shear wave propagation, and shear wave interactions in a medium.

As illustrated in figure 2(a), the simulated model consisted of a stiff branching y-shaped tube placed within a uniform background material. The model was represented as a cube with dimensions of 120 mm × 120 mm × 120 mm, and the background material was assumed to be a uniform isotropic material with a SWS of 1 m/s. The y-shaped tube was also assumed to be a uniform isotropic material with a cross-section radius of 15 mm and a SWS of 3 m/s. In the simulation, the viscoelastic properties of the materials were defined using the classical Kelvin-Voigt absorption model. This model considers the absorption coefficient to be proportional to the square of the frequency in the low-frequency limit (Treeby and Cox 2010a, 2010b). For both the background material and the y-shaped tube, the absorption coefficient was set to 0.5 $dB/(MHz^2\ cm)$. By incorporating the viscoelastic properties and absorption coefficients based on the Kelvin-Voigt model, the simulation is designed to capture the realistic behavior of shear wave propagation and absorption in the simulated materials.

The shear waves were generated by the excitation sources placed in random positions within the simulated medium. This approach ensured the generation of a fully reverberant shear wave field. To verify that the excitation sources did not directly generate waves within the y-shaped tube, the length of the tube was made smaller than the overall dimensions of the cube. This arrangement allowed the waves to be generated solely within the cube and then propagate both inside the cube and the tube. This setup reflects the typical scenario in real elastography experiments, where the waves are generated outside the region of interest (ROI) and propagate into the target area.

A white Gaussian noise with a signal-to-noise ratio (SNR) of 10 dB was introduced into the shear wave field in the k-Wave simulation. The addition of this noise in the simulation enables the evaluation of the robustness and performance of the AIA approach under more realistic conditions. The impact of noise on the accuracy performance of AIA is further examined in the Appendix.

Furthermore it was assumed that in this elastography experiment only the velocity field along the *z*-axis could be measured. This assumption aligns with imaging modalities where the sensitive displacement measurements are mainly directed along one specific direction. By including these considerations in the simulation, a realistic experimental setup that closely resembles the conditions encountered in practical elastography studies was created. The particle velocity field along the *z*-axis, after sufficient time for complete wave interaction and the generation of a reverberant shear wave field, is shown in figure 2(b). Figure 2(c) presents a focused view of the particle velocity field within the y-shaped tube. It is clear in this figure that a fully reverberant shear wave field is generated in both the background and the y-shaped tube. The distinct difference between the wave amplitudes and wavelengths in a y-shaped tube and the background makes the y-shaped tube recognizable in the shear wave field. To reduce computational costs, the 3D ROI has been selectively cut.

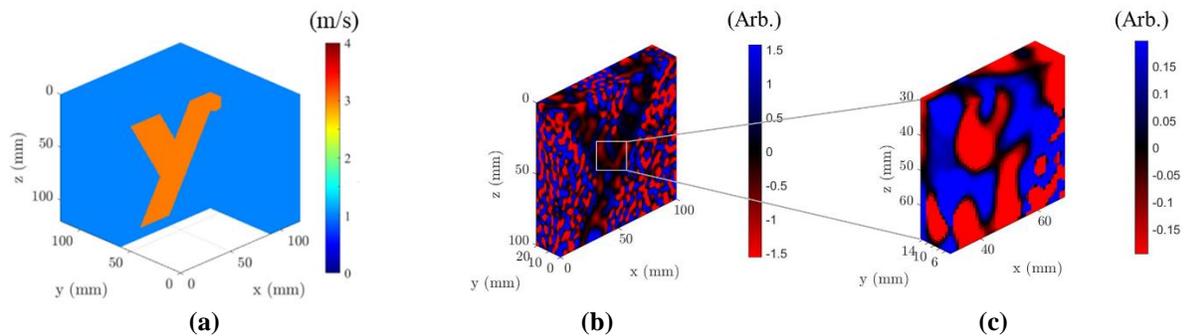

(a)      (b)      (c)

**Figure 2.** k-Wave elastography simulation: (a) SWS properties in the k-Wave simulation including a branching y-shaped tube in a uniform background (b) particle velocity field within the background material and y-shaped tube along the *z*-axis at a frequency of 200 Hz, (c) a focused view of particle velocity field within the y-shaped tube.

### 3.2. Ultrasound elastography

In this section, the effectiveness of the AIA estimator for estimating the SWS in ultrasound elastography is investigated. To provide a comprehensive assessment across various experimental scenarios and conditions, two different ultrasound experiments with different excitation systems and field of view configurations were studied. In the first study, a series of ultrasound elastography experiments were performed utilizing an elastic breast phantom containing a stiff lesion. These experiments were conducted at different excitation frequencies. In these experiments, imperfect reverberant shear wave fields were generated using two excitation sources positioned on opposite sides of the phantom. The second study focused on the utilization of datasets obtained from *in vivo* ultrasound elastography of the human liver-kidney region with a fully reverberant shear wave field. In this case, data were collected at different excitation frequencies.

*3.2.1. Breast phantom ultrasound elastography*
In the first ultrasound elastography study, a CIRS breast phantom (model 509, CIRS Inc., Norfolk, Virginia, USA) was utilized. The CIRS breast phantom is designed to simulate human breast tissue characteristics and closely replicates the mechanical properties of actual breast tissue. The breast phantom contains several lesions of different sizes. In this study, to illustrate the capability of the AIA estimator to differentiate the stiffer regions from the background material, the focus was directed toward a specific 10 mm diameter lesion within the phantom. To generate a reverberant shear wave field, two miniature vibration sources (model NCM02-05-005-4 JB, H2W, Linear Actuator, Santa Clara, CA, USA) were employed in contact with the breast phantom. The excitation signal for the vibration sources was provided by a power amplifier (model 2718, Bruel and Kjaer, Naerum, Denmark), and a digital power amplifier (model LP-2020 A+, Lepai, Bukang, China) driven by a dual channel function generator (model AFG3022B, Tektronix, Beaverton, OR, USA). The ultrasound elastography experiment was performed at various excitation frequencies including 900 Hz, 600 Hz, and 400 Hz.
The ultrasound elastography experiment utilized a Verasonics ultrasound system (V-1, Verasonics, Kirkland, WA, USA). This system enables high frame rate acquisition and a coherent plane wave compounding acquisition scheme. The Verasonics system was connected to a linear array ultrasound transducer (Model L7-4, ATL, Bothell, WA, USA). To track the induced displacements caused by shear wave propagation, a Loupas estimator was employed (Loupas *et al* 1995). In the ultrasound elastography experiment, a 3D matrix of in-phase and quadrature (IQ) data was collected and stored for subsequent postprocessing. The IQ data contains information about the amplitude and phase of the ultrasound signals received by the transducer. To calculate the axial particle displacements, frame-to-frame analysis was performed on the acquired 3D IQ data. Then, the amplitude and phase of displacement at each pixel were estimated. In all the experiments, the center frequency of the ultrasound transducer was set to 5 MHz. The tracking pulse repetition frequency (PRF) was adjusted to acquire at least 20 samples per cycle of the vibration frequency. Further information regarding the characteristics of the phantom, the parameters of the ultrasound elastography experiment, and data postprocessing to determine the displacement field can be found in Ormachea and Parker (2021).
Figure 3(a) displays the B-mode ultrasound scan of the breast phantom with a 10 mm diameter lesion positioned at the center. In this experimental data, shear waves were generated within the phantom at various excitation frequencies, and ultrasound elastography scans were performed during the excitation process. The phase map of the shear wave field at the frequency of 900 Hz is presented in figure 3(b). This representation describes an imperfect reverberant shear wave field within the phantom that is more bi-directional. As evident in the figure, the lesion is visible in the B-mode scan while it cannot be recognized in the shear wave field.

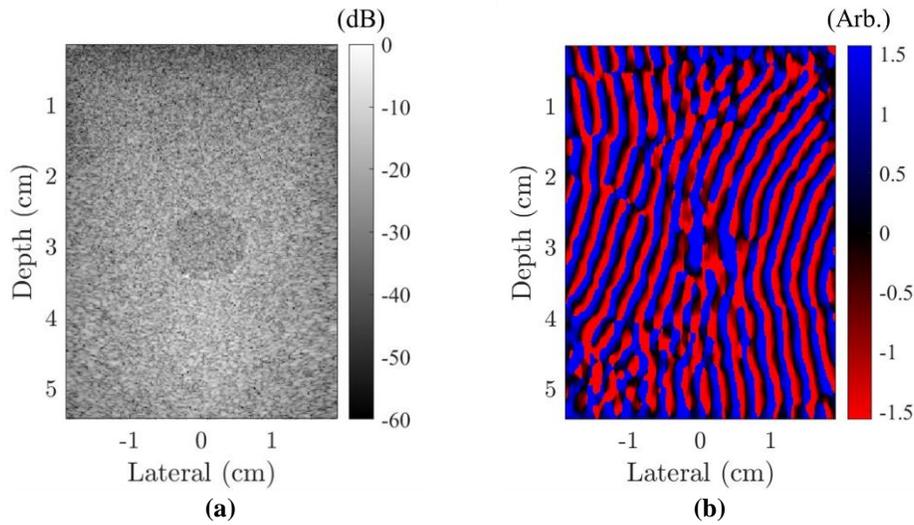

**Figure 3.** Ultrasound elastography of a breast phantom with a lesion at the center: (a) B-mode ultrasound scan of the phantom, (b) phase map of the shear wave field at the frequency of 900 Hz.

*3.2.2. Liver-kidney ultrasound elastography*
In the next study to assess the effectiveness of the AIA estimator, datasets derived from ultrasound elastography of the human liver-kidney region were investigated. The liver scan was obtained in conjunction with the small study reported in Ormachea *et al* (2019) which was performed under the requirements of informed consent of the Southwoods Imaging Clinical Institutional Review Board. A Verasonics ultrasound system (Vantage-128™, Verasonics, Kirkland, WA, USA), connected to a convex ultrasound probe (model C4-2, ATL, Bothell, WA, USA) was used with a 3 MHz center frequency. These datasets include the B-mode ultrasound scan and displacement field data from fully reverberant shear waves at a broad range of excitation frequencies including 702 Hz, 585 Hz, 468 Hz, 351 Hz, 234 Hz, and 117 Hz. Additional details about the data postprocessing to obtain the displacement field are provided in Ormachea *et al* (2019).

Figure 4(a) displays the B-mode ultrasound scan, which provides an anatomical view of the liver and kidney. In the scan, the different tissue layers can be observed, including the abdominal muscle and fat layers (zone A), the distinct darker zones representing the liver (zone L), and the kidney (zone K). Figure 4(b) presents the phase map of the shear wave field at the frequency of 702 Hz. This representation depicts a randomized, reverberant shear wave field, throughout the abdominal region.

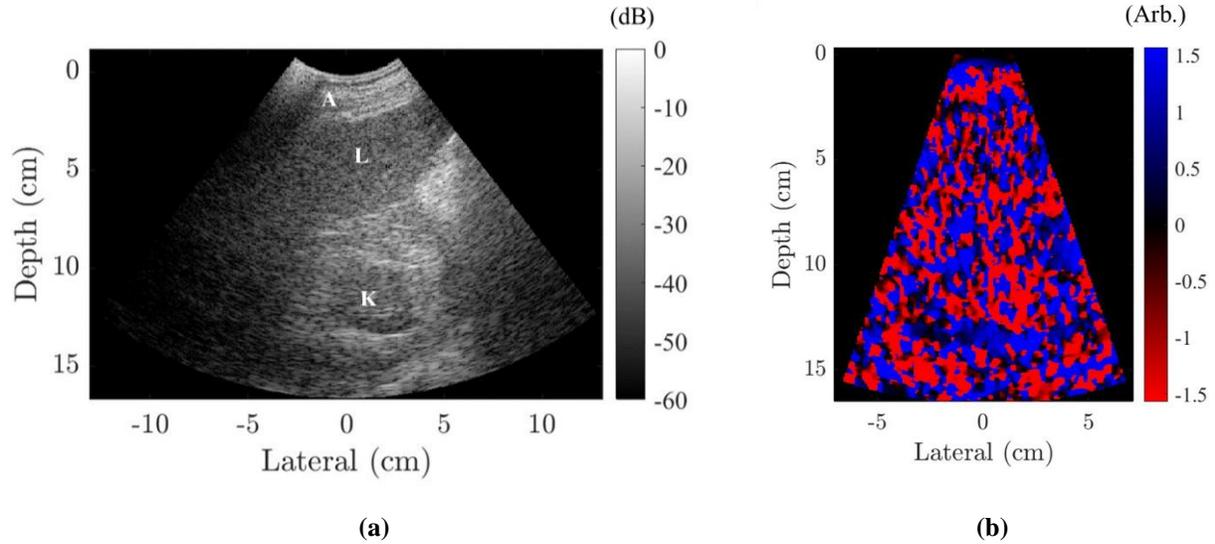

**Figure 4.** Ultrasound elastography of a human liver-kidney: (a) B-mode ultrasound scan of the human liver-kidney region including the abdominal layer (zone A), the liver (zone L), and the kidney (zone K), (b) phase map of the reverberant shear wave field at the frequency of 702 Hz.

### 3.3. Magnetic resonance elastography

In the next step, an MRE experiment was conducted on a brain-mimicking phantom with two stiff lesions. Figure 5(a) displays the brain-shaped phantom, along with the coordinate system for reference. The brain-shaped phantom was constructed using a suspension composed of bovine gelatin (200 bloom: Sigma Aldrich Chemicals, St. Louis, MO, USA), de-ionized water (18 MΩ), and ethylenediamine tetra-acetic acid (Sigma Aldrich Chemicals, St. Louis, MO, USA). A controlled and reproducible process was used (as described by Doyley *et al* (2003)) to fabricate the phantom with precise dimensions of 180 mm in the long axis, 130 mm in the short axis, and 70 mm in height. The brain phantom contained two spherical gelatin lesions, measuring 18 mm and 12 mm in diameter, respectively. The gelatin concentration in the background region was 8%, while both lesions had a gelatin concentration of 18%. The remaining composition for the background consisted of 92% water, while the lesion composition consisted of 81.64% water and 0.36% copper sulfate. The addition of copper sulfate served to provide contrast between the lesion and the background, enhancing the anatomical image in the scan. The fabrication process ensured the creation of a brain-shaped phantom with accurately defined dimensions and distinct gelatin concentrations, allowing for reliable experimental testing and imaging analysis.

The shear wave field was generated within the phantom at the frequency of 200 Hz by a pneumatic mechanical actuator with a passive driver (Resoundant, Inc., Rochester, MN, USA) which moved in the *y*-direction. All elastographic imaging procedures were conducted using a whole-body 3T MRI scanner (Prisma, Siemens, Erlangen, Germany) equipped with a 20-channel head coil. Each 3D data set acquisition required approximately six minutes. Data collection in MRE was carried out with a 1.6 mm isotropic voxel size. The single-shot echo-planar imaging (EPI) sequence (Johnson *et al* 2014) was utilized to measure the resulting time-varying harmonic tissue displacements. More details about the phantom properties, MRE experiment parameters, and data postprocessing to determine displacement field are described in Kabir *et al* (2023).

In the MRI scan of the brain phantom (see figure 5(b)), two spherical lesions are clearly visible. The phase map of X-motion and Z-motion displacement fields at the frequency of 200 Hz are displayed in figures 5(c) and (d), respectively. It is worth noting that X-motion and Z-motion displacement fields were exclusively employed as they exhibited the strongest signal. As shown in figure 5, shear wave fields with the dominant directions are generated in both X-motion and Z-motion. The presence of lesions is visible within these displacement fields as longer spatial wavelengths and the bending of wavefronts around the lesions.

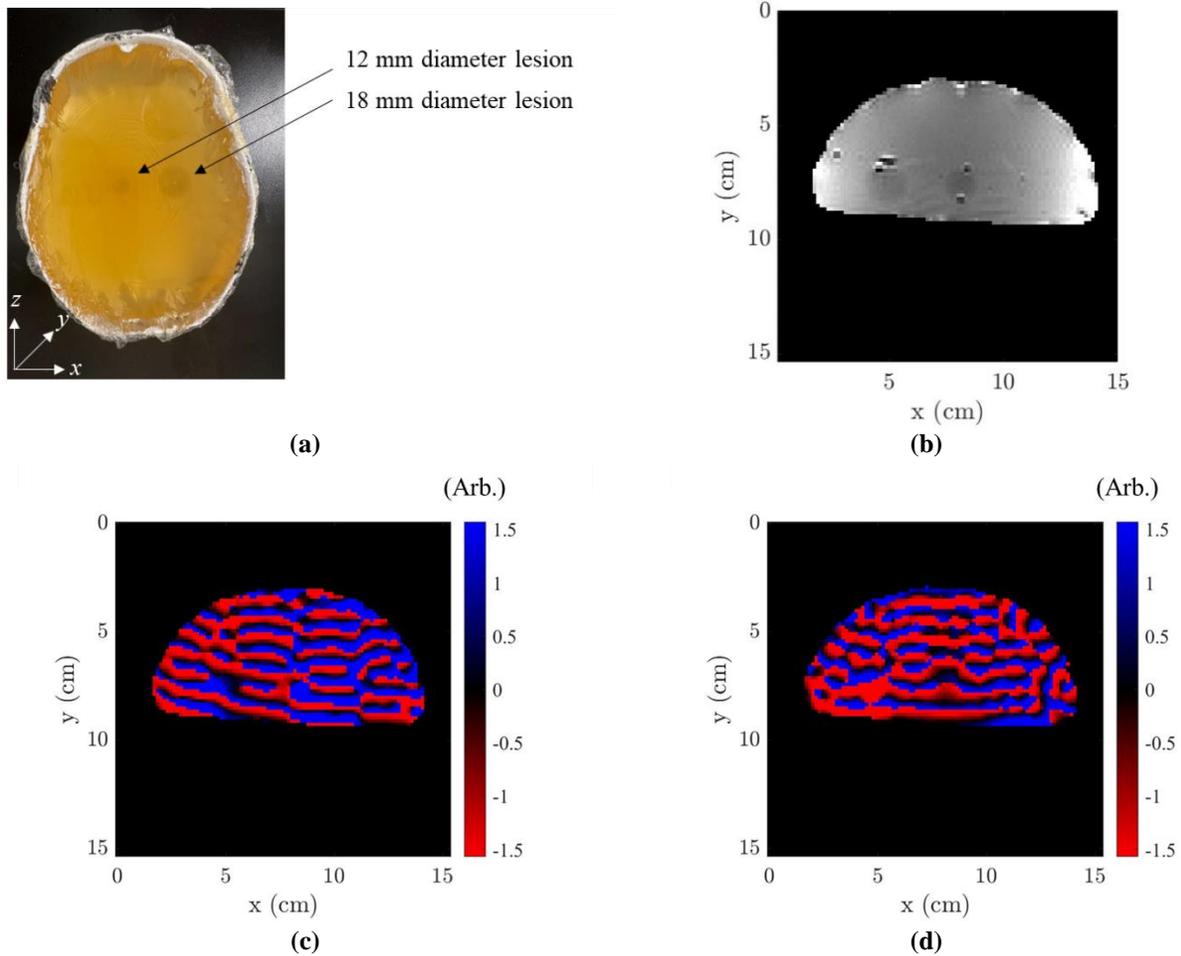

**Figure 5.** MRE of a brain phantom with two lesions: (a) brain phantom shape and the coordinate system for MRE, (b) MRI scan of the brain phantom (c) phase map of X-motion, and (d) phase map of Z-motion displacement fields at the frequency of 200 Hz.

### 3.4. Optical coherence elastography

To assess the effectiveness of the AIA approach in OCE, an OCE experiment was conducted on a gelatin phantom featuring a cylindrical stiff lesion. For this experiment, a custom-built OCT system equipped with a swept-source laser with a center wavelength of 1310 nm and a bandwidth of 140 nm (HSL-2100-HW, Santec, Aichi, Japan) was employed. The lateral and axial resolutions of the system in the air were 20 $\mu$m and 6 $\mu$m, respectively, with a maximum sensitivity of approximately 110 dB. The field of view for this experiment was adjusted to be 10 mm × 10 mm. Synchronization between the swept-source optical coherence tomography (SS-OCT) and a mechanical excitation system was achieved using LabVIEW (Version 14, National Instruments, Austin, Texas, USA).

The phantom and its lesion were composed of 5% gelatin concentration and 10% gelatin powder concentration respectively, designed to exhibit distinct mechanical properties and SWS characteristics. To introduce the light scattering properties to the phantom, a 3% concentration of intralipid powder (coffee creamer) was added to both the background and lesion. The remaining composition for the background consisted of 1% salt and 91% water, while the lesion composition comprised 1% salt and 86% water.

A reverberant shear wave field at the frequency of 1500 Hz was generated within the phantom using a multi-pronged ring actuator connected to a piezoelectric device. Additional information regarding the optical setup and excitation system used in OCE can be found in Ge *et al* (2022b). The acquisition approach and data postprocessing developed by Zvietcovich *et al* (2019b) were employed to obtain the displacement field.

A three-dimensional OCT scan along with the corresponding displacement field of shear waves for the phantom containing a stiff lesion at the frequency of 1500 Hz are presented in figure 6. Notably, a pie-cut is applied to view the interior. In the three-dimensional OCT scan (see figure 6(a)), the presence of the cylindrical lesion is visually revealed through the height difference between the phantom and the lesion at the center. However, it should be emphasized that both the phantom and its lesion were intentionally designed with identical scattering properties. Therefore, despite differences in surface height, the 3D OCT scan does not distinguish between the stiff lesion and background due to their matching scattering properties. Figure 6(b) clearly illustrates the generation of circular waves at the boundaries of the phantom, which subsequently propagate and interfere with each other within the internal region. This results in a wave field that exhibits a more directional nature at the outer phantom boundaries and a more reverberant behavior within the internal region.

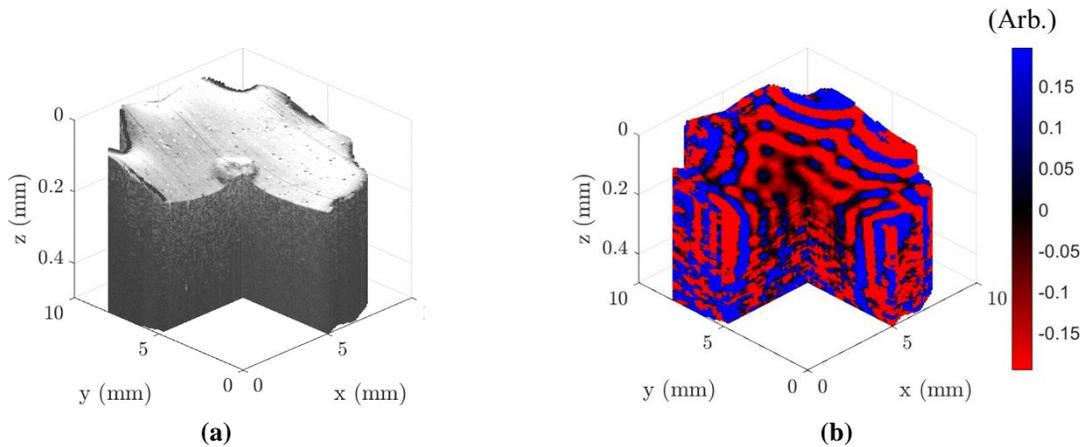

**Figure 6.** OCE of a phantom containing a lesion: (a) 3D OCT scan of the phantom with a pie-cut, (b) displacement field of shear waves at the frequency of 1500 Hz.

## 4. Results and discussion
### 4.1. Numerical simulation

In the k-Wave simulation, access to the complete 3D dataset, allowed for the analysis of shear wave behavior in various planes, including the *xy* plane, *xz* plane, and *yz* plane. Within each plane, measurements were conducted for the simple baseline autocorrelation functions in two directions and the AIA function. Specifically, for the *xy* plane, measurements involved the simple baseline autocorrelation function along the *x*-axis, denoted as $B_{V_zV_z}(\Delta\varepsilon_x, \Delta t)$ and the *y*-axis, denoted as $B_{V_zV_z}(\Delta\varepsilon_y, \Delta t)$, as well as the AIA in the *xy* plane, denoted as $B_{AI_{xy}}(\Delta\rho, \Delta t)$. Since we assumed that the detected velocity field is aligned with the *z*-axis, the presumed imaging direction, and the autocorrelation function in the *xy* plane remains consistent regardless of the autocorrelation direction. In other words, the simple baseline autocorrelation functions yield similar results for both $B_{V_zV_z}(\Delta\varepsilon_x, \Delta t)$ and $B_{V_zV_z}(\Delta\varepsilon_y, \Delta t)$ directions, and $\theta_s$ in equation (5) is equal to $\pi/2$ for both $\Delta\varepsilon_x$ and $\Delta\varepsilon_y$, resulting in equations (6.a) and (6.b). Similarly, measurements of the simple baseline autocorrelation function were conducted along the *x*-axis $B_{V_zV_z}(\Delta\varepsilon_x, \Delta t)$ and the *z*-axis $B_{V_zV_z}(\Delta\varepsilon_z, \Delta t)$, in the *xz* plane as well as along the *y*-axis $B_{V_zV_z}(\Delta\varepsilon_y, \Delta t)$ and the *z*-axis $B_{V_zV_z}(\Delta\varepsilon_z, \Delta t)$, in the *yz* plane. In both the *xz* and *yz* planes, the angle $\theta_s$ for $\Delta\varepsilon_x$ and $\Delta\varepsilon_y$ is equal $\pi/2$ and the simple baseline autocorrelation functions follow equations (6.a) and (6.b). For $\Delta\varepsilon_z$, $\theta_s$ is zero and the simple autocorrelation function $B_{V_zV_z}(\Delta\varepsilon_z, \Delta t)$ follows equation (6.c). It is important to note that when measuring simple baseline autocorrelation functions in other directions with different $\theta_s$, equation (5) should be employed. Through the AIA estimation conducted in each *xy*, *xz*, and *yz* plane (per equations (8) and (9)) a comprehensive analysis of the shear waves in different planes was obtained, providing valuable insights for SWS estimation and tissue characterization. Figure 7 presents the SWS measured from the k-Wave simulation at a frequency of 200 Hz using the simple baseline autocorrelation approaches in different planes and directions (left column), and our proposed AIA approach in different planes (middle column). The autocorrelation window size in all measurements is set as 15 mm × 15 mm.

Table 1 showcases a comparison of the estimated SWS using simple autocorrelation approaches and the AIA approach in the background material and the branching tube across different autocorrelation planes. It is worth noting that the SWS and SNR for all measurements were computed using the almost full 3D background and y-shaped tube with some distance to the tube-background boundaries. The average error for estimating background SWS was 7% using either simple autocorrelation approaches or AIA. However, the average error for estimating SWS in the branching tube using simple autocorrelation approaches was 9%, and using the AIA approach was 3%. Another important parameter in elastography estimation is the ratio of tube SWS to background SWS.

Table 2 presents a comparison between the estimated SWS ratio using simple autocorrelation approaches and the AIA approach in different autocorrelation planes. The average error of the estimated SWS ratio using simple autocorrelation approaches is 14% whereas using the AIA approach reduces this error to 4%. Table 3 presents the SNR of estimated SWS in the background material for k-Wave simulation using simple autocorrelation approaches and the AIA approach along with the improvement achieved by the AIA approach in different autocorrelation planes (*xy*, *xz*, and *yz* planes).

The results clearly demonstrate that the AIA approach significantly enhances the SNR in estimating SWS in the background with an average improvement of 28% at the frequency of 200 Hz. Moreover, we can further improve the estimation by calculating the median, or averaging the estimated SWS in the *xy*, *xz*, and *yz* autocorrelation planes, as shown in the right column of figure 7. The application of averaging leads to an SNR increase of more than 43%, while employing the median increases the SNR by more than 45% at the excitation frequency of 200 Hz. These findings highlight the effectiveness of the AIA approach and its potential to enhance the accuracy and reliability of SWS estimation.

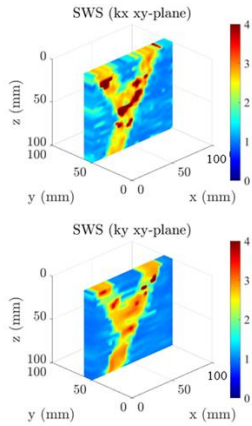
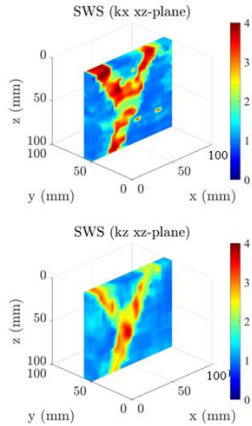
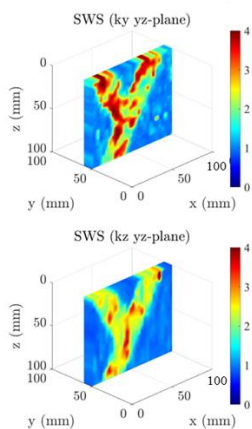
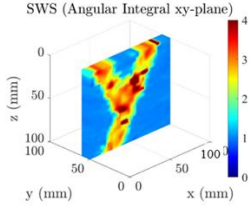
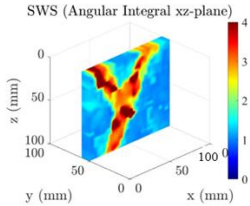
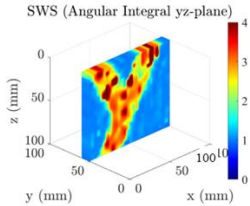
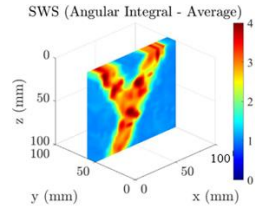
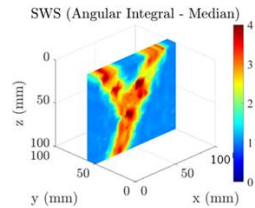

**Figure 7.** SWS measured from the k-Wave simulation at the frequency of 200 Hz using the simple baseline autocorrelation approaches in different planes and directions (left column), using the AIA approach in different autocorrelation planes (middle column), the average and median SWS measured by AIA approach in different autocorrelation planes (right column).

**Table 1.** SWS in the background material and the branching tube estimated using simple autocorrelation approaches and the AIA approach in k-Wave simulation across different autocorrelation planes.

| Autocorrelation plane | SWS (m/s) in background using simple autocorrelation (first axis) | SWS (m/s) in background using simple autocorrelation (second axis) | SWS (m/s) in background using AIA | SWS (m/s) in branching tube using simple autocorrelation (first axis) | SWS (m/s) in branching tube using simple autocorrelation (second axis) | SWS (m/s) in branching tube using AIA |
|---|---|---|---|---|---|---|
| *xy* plane | 1.12 ± 0.14 | 1.03 ± 0.09 | 1.02 ± 0.04 | 2.90 ± 0.74 | 2.70 ± 0.42 | 2.98 ± 0.57 |
| *xz* plane | 1.08 ± 0.23 | 1.08 ± 0.13 | 1.14 ± 0.12 | 2.93 ± 0.61 | 2.51 ± 0.47 | 3.19 ± 0.74 |
| *yz* plane | 1.07 ± 0.17 | 0.97 ± 0.08 | 1.05 ± 0.08 | 2.81 ± 0.68 | 2.50 ± 0.36 | 3.08 ± 0.65 |

**Table 2.** SWS ratio between the branching tube and the background material estimated using simple autocorrelation approaches and the AIA approach in k-Wave simulation across different autocorrelation planes.

| Autocorrelation plane | SWS ratio using simple autocorrelation (first axis) | SWS ratio using simple autocorrelation (second axis) | SWS ratio using AIA | SWS ratio estimation error using simple autocorrelation (first axis) | SWS ratio estimation error using simple autocorrelation (second axis) | SWS ratio estimation error using AIA |
|---|---|---|---|---|---|---|
| *xy* plane | 2.59 ± 0.74 | 2.62 ± 0.47 | 2.92 ± 0.57 | 14% | 13% | 3% |
| *xz* plane | 2.71 ± 0.81 | 2.32 ± 0.52 | 2.80 ± 0.71 | 10% | 23% | 7% |
| *yz* plane | 2.63 ± 0.76 | 2.58 ± 0.43 | 2.93 ± 0.66 | 12% | 14% | 2% |

**Table 3.** SNR of estimated SWS in the background material for k-Wave simulation using simple autocorrelation approaches and the AIA approach across different autocorrelation planes.

| Autocorrelation plane | SNR (dB) in background using simple autocorrelation (first axis) | SNR (dB) in background using simple autocorrelation (second axis) | SNR (dB) in background using AIA | AIA improvement over simple autocorrelation (first axis) | AIA improvement over simple autocorrelation (second axis) |
|---|---|---|---|---|---|
| *xy* plane | 8.94 | 10.61 | 13.69 | 53% | 29% |
| *xz* plane | 6.78 | 9.13 | 9.68 | 43% | 6% |
| *yz* plane | 8.12 | 10.69 | 10.98 | 35% | 3% |

Extensive comparisons of the SWS ratio and SNR using the average of simple autocorrelation approaches and the AIA approach across different autocorrelation planes are explored in table A.1 and table A.2 in the Appendix. Furthermore, the effect of different levels of added noise on the SWS estimated using the AIA approach across different autocorrelation planes is investigated through figures A.2 and A.3 and table A.3 in the Appendix.

### 4.2. Ultrasound elastography

In ultrasound elastography, the dataset primarily comprises the *z*-oriented displacement field within the imaged *xz* plane. This data reflects the propagation and interaction of shear waves within the examined tissue. To estimate the SWS map, a simple autocorrelation approach in the *x*-direction, a simple autocorrelation approach in the *z*-direction, and the AIA approach were employed.

#### 4.2.1. Breast phantom ultrasound elastography

In our study, different autocorrelation approaches were applied to estimate SWS in ultrasound elastography of the breast phantom with a lesion, conducted under conditions of imperfect reverberant shear wave fields at various excitation frequencies, including 400 Hz, 600 Hz, and

900 Hz. For these measurements, a square autocorrelation window of dimensions 7.5 mm was utilized.

Figure 8 displays the estimated SWS using simple baseline autocorrelation approaches in the *x*-direction (see figure 8(a)) and *z*-direction (see figure 8(b)), as well as the AIA approach (see figure 8(c)) for the excitation frequency of 900 Hz. As shown in figure 3(b), the shear wave field exhibits an imperfect reverberant nature with prominently directional waves in the *x*-direction. Under these circumstances, the baseline autocorrelation in the *z*-direction measures a longer wavelength, resulting in an overestimation of SWS as evident in figure 8(b). Consequently, the baseline autocorrelation in the *z*-direction does not yield satisfactory results for SWS estimation, even the difference in SWS between the phantom and lesion is undetectable as seen in figure 8(b).

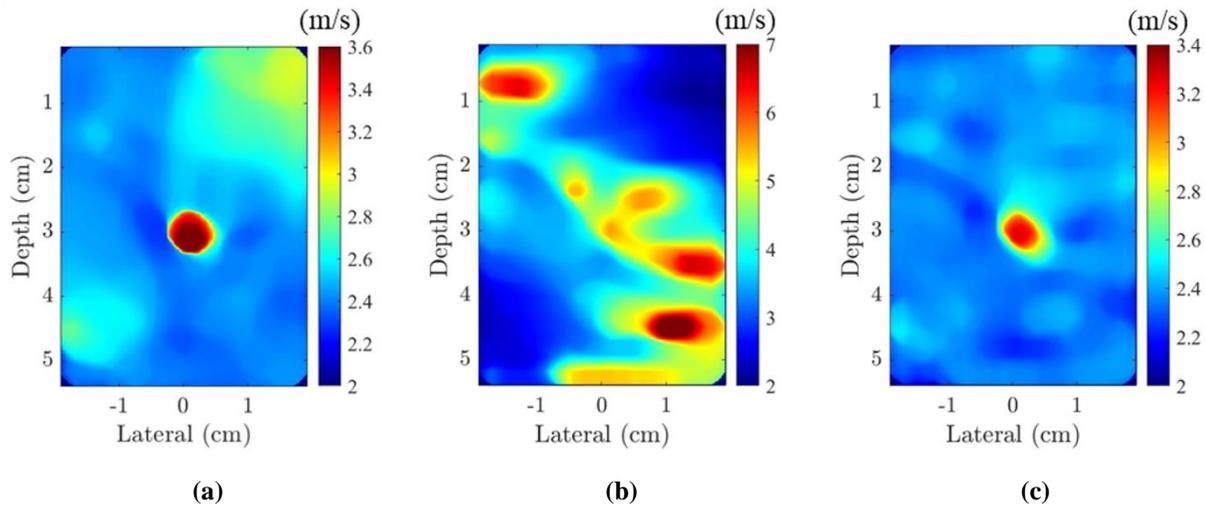

**Figure 8.** Estimated SWS using (a) simple autocorrelation in *x*-direction (b) simple autocorrelation in *z*-direction, and (c) the AIA approach for ultrasound elastography of a breast phantom with a lesion at the frequency of 900 Hz.

The mean SWS and the standard deviation for the CIRS breast phantom were reported as 2.27 ± 0.19 m/s in Ormachea and Parker (2021). Table 4 presents the estimated SWS in the background using simple autocorrelation approaches ($k_x$ and $k_z$) and the AIA approach across different excitation frequencies. The average SWS and SNR for simple autocorrelation approaches and AIA were calculated using the almost full 2D background with some distance to the lesion boundaries. The average SWS estimated using simple autocorrelation $k_x$ (parallel to the wave direction) and AIA at different excitation frequencies closely align with the reported value by Ormachea and Parker (2021). However, as anticipated, the simple autocorrelation $k_z$ (perpendicular to the wave direction) estimates notably higher SWSs for the background at different excitation frequencies. Considering the reported value of 2.27 ± 0.19 m/s as the ground truth, the SWS estimation error for simple autocorrelation approaches and AIA are calculated and presented in table 4. The average error for SWS estimation using AIA is 6% while it is 6% and 53% using $k_x$ and $k_z$ respectively.

**Table 4.** SWS in the background material for ultrasound elastography of the breast phantom estimated using simple autocorrelation approaches ($k_x$ and $k_z$) and the AIA approach across different excitation frequencies, errors were calculated using the SWS value of 2.27 (m/s) for the breast phantom as obtained in Ormachea and Parker (2021).

| Excitation frequency | SWS (m/s) in background using simple autocorrelation ($kx$) | SWS (m/s) in background using simple autocorrelation ($kz$) | SWS (m/s) in background using AIA | SWS estimation error using simple autocorrelation ($kx$) | SWS estimation error using simple autocorrelation ($kz$) | SWS estimation error using AIA |
|---|---|---|---|---|---|---|
| 900 Hz | 2.49 ± 0.17 | 3.74 ± 1.09 | 2.45 ± 0.11 | 10% | 65% | 8% |
| 600 Hz | 2.34 ± 0.13 | 3.36 ± 0.91 | 2.22 ± 0.11 | 3% | 48% | 2% |
| 400 Hz | 2.17 ± 0.13 | 3.32 ± 0.75 | 2.10 ± 0.12 | 4% | 46% | 7% |

Table 5 presents the SNR of estimated SWS within the background material for ultrasound elastography of the breast phantom, employing simple autocorrelation approaches ($kx$ and $kz$) as well as the AIA approach at different excitation frequencies. The estimated SWS using AIA demonstrates a major 150% improvement in SNR within the background region when compared to the $kz$ approach. The SWS obtained using the AIA approach also displays an enhancement in estimation accuracy compared to the simple autocorrelation approach in the *x*-direction ($kx$). This improvement is evidenced by a 14% increase in the SNR within the background region. Similar outcomes were obtained at excitation frequencies of 600 Hz and 400 Hz. The presented results in Figure 8 and both table 4 and table 5 indicate that AIA is effective in estimating SWS even in the presence of a highly directional wave field.

**Table 5.** SNR of estimated SWS in the background material for the breast phantom ultrasound elastography using simple autocorrelation approaches and the AIA approach across different excitation frequencies.

| Excitation frequency | SNR (dB) in background using simple autocorrelation ($kx$) | SNR (dB) in background using simple autocorrelation ($kz$) | SNR (dB) in background using AIA | AIA improvement over $kx$ | AIA improvement over $kz$ |
|---|---|---|---|---|---|
| 900 Hz | 11.73 | 5.36 | 13.38 | 14% | 150% |
| 600 Hz | 12.49 | 5.66 | 13.01 | 4% | 130% |
| 400 Hz | 12.35 | 6.43 | 12.44 | 1% | 93% |

*4.2.2. Liver-kidney ultrasound elastography*

Figures 9(a) and (b) present the SWS obtained from simple baseline autocorrelation functions in the *x*-direction and *z*-direction at the frequency of 702 Hz, while figure 9(c) illustrates the SWS obtained using the AIA approach at the frequency of 702 Hz. All measurements depicted in figure 9 were conducted using a square autocorrelation window of dimension 18.5 mm. To evaluate the accuracy of the AIA approach, the SWS elastogram at the frequency of 702 Hz is associated with the ultrasound B-mode scan in figure 9(d). As evident, the estimated SWS using the AIA estimator effectively outlined the structural details of tissues within the liver-kidney region including the abdominal muscle and fat layers (zone A), the liver (zone L), and the kidney (zone K), resembling the structures observed in the B-mode scan.

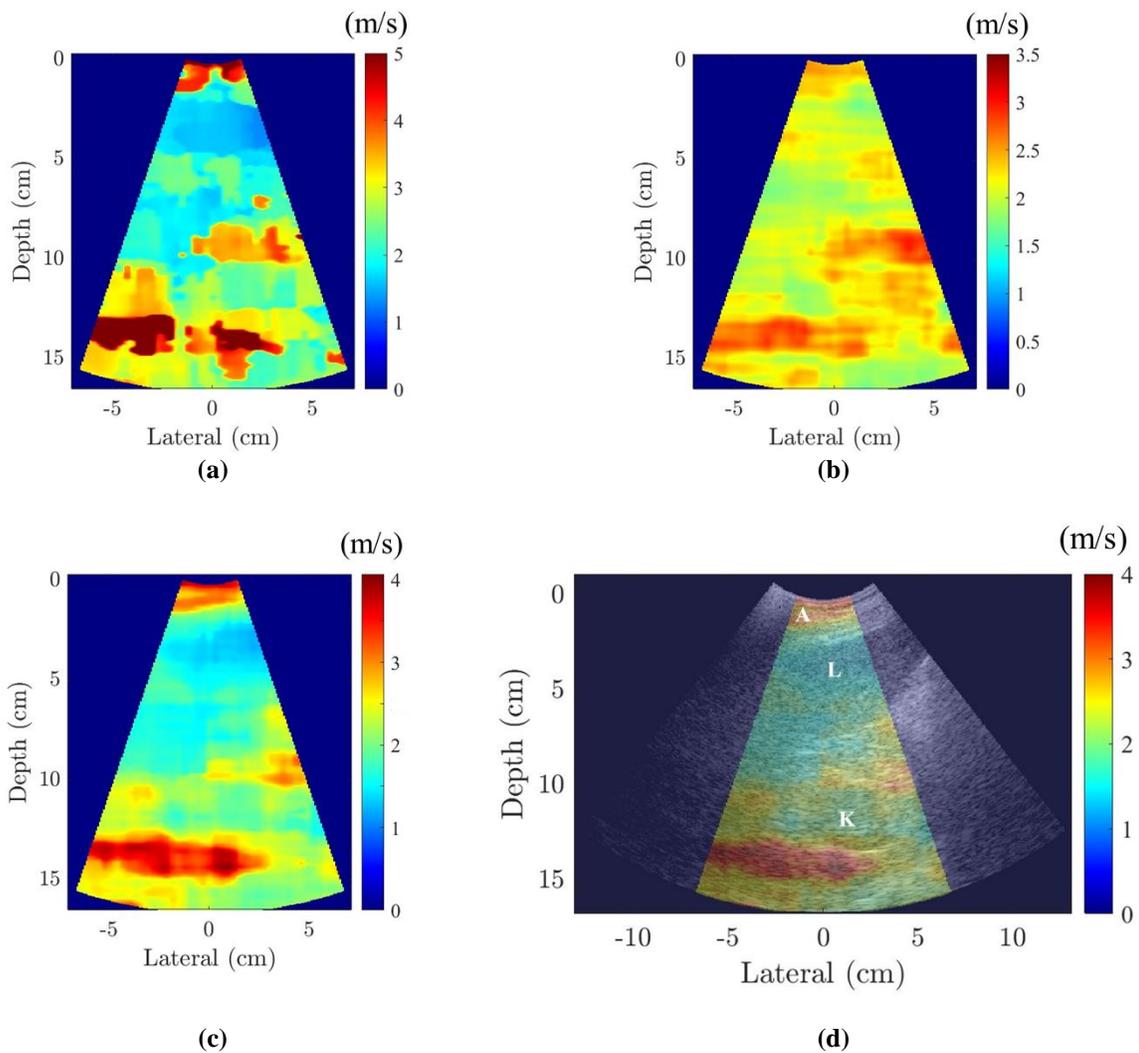

**Figure 9.** Estimated SWS of the liver-kidney ultrasound elastography at the frequency of 702 Hz using (a) simple autocorrelation in *x*-direction (b) simple autocorrelation in *z*-direction, (c) the AIA approach, (d) SWS map obtained from the AIA approach overlaid on the ultrasound B-mode scan of the liver-kidney region highlighting abdominal layer (zone A), the liver (zone L), and the kidney (zone K).

Consistent outcomes were achieved across a range of excitation frequencies, including 585 Hz, 468 Hz, 351 Hz, 234 Hz, and 117 Hz. In order to analyze the data and compute the SWS and SNR, the largest region of interest (ROI) that can be selected within the liver and kidney were considered for simple autocorrelation approaches and AIA. The frequency resolved mean and standard deviation values for the liver and the kidney are presented in Table 6.

**Table 6.** SWS in the liver and the kidney estimated using simple autocorrelation approaches and the AIA approach in liver-kidney ultrasound elastography across different excitation frequencies.

| Excitation frequency | SWS (m/s) in liver using simple autocorrelation ($k_x$) | SWS (m/s) in liver using simple autocorrelation ($k_z$) | SWS (m/s) in liver using AIA | SWS (m/s) in kidney using simple autocorrelation ($k_x$) | SWS (m/s) in kidney using simple autocorrelation ($k_z$) | SWS (m/s) in kidney using AIA |
|---|---|---|---|---|---|---|
| 702 Hz | 3.13 ± 0.12 | 4.07 ± 0.12 | 2.69 ± 0.08 | 4.93 ± 0.42 | 4.47 ± 0.25 | 3.98 ± 0.22 |
| 585 Hz | 3.12 ± 0.54 | 3.12 ± 0.15 | 2.34 ± 0.25 | 3.81 ± 0.37 | 3.30 ± 0.37 | 2.98 ± 0.14 |
| 468 Hz | 2.71 ± 0.27 | 2.72 ± 0.11 | 1.92 ± 0.20 | 3.11 ± 0.18 | 2.56 ± 0.10 | 2.48 ± 0.10 |
| 351 Hz | 2.47 ± 0.20 | 2.05 ± 0.12 | 1.96 ± 0.11 | 3.25 ± 0.90 | 2.42 ± 0.27 | 2.60 ± 0.28 |
| 234 Hz | 1.61 ± 0.14 | 2.02 ± 0.10 | 1.60 ± 0.09 | 2.18 ± 0.39 | 1.37 ± 0.05 | 1.66 ± 0.09 |
| 117 Hz | 0.80 ± 0.28 | 0.80 ± 0.05 | 0.71 ± 0.08 | 0.95 ± 0.29 | 0.89 ± 0.05 | 0.88 ± 0.08 |

Figures 10(a) and (b) represent the SNR of measured SWS, employing simple autocorrelation approaches ($k_x$ and $k_z$) and the AIA approach within the liver and kidney regions, respectively. The results illustrate that the SNR for simple autocorrelation in the *z*-direction ($k_z$) is generally higher than the SNR for AIA. However, an examination of the estimated SWS ratio between the liver and kidney in figure 10(c) reveals that the SWS ratio estimated by $k_z$ consistently remains around 1.0 at different excitation frequencies. This indicates that $k_z$ consistently estimates nearly the same SWS for both the liver and kidney, which is inaccurate, as the kidney is commonly found to be stiffer than the liver (Lee *et al* 2013, Johnson *et al* 2021). Furthermore, as evident in figure 9(b), the estimated SWS using $k_z$ fails to visualize the structural details of tissues within the liver-kidney region.

These results emphasize the effectiveness of the AIA approach in accurately estimating and visualizing the SWS in ultrasound elastography of the *in vivo* liver-kidney region. Consequently, by incorporating the AIA approach, we have the potential to extend the diagnostic capabilities of ultrasound elastography, providing valuable insights for tissue characterization and differentiation of tissue properties.

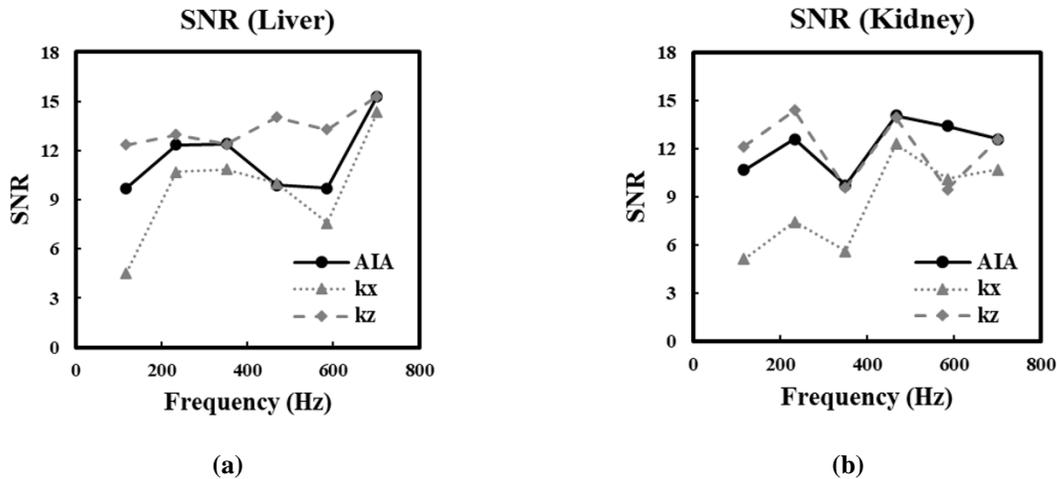

(a)      (b)

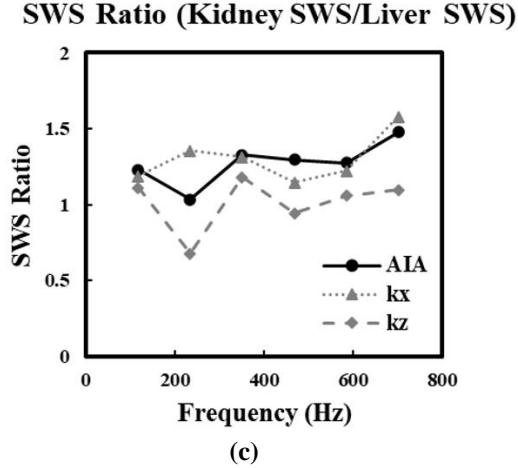

(c)

**Figure 10.** SNR of SWS measured using simple autocorrelation approaches and the AIA approach at different excitation frequencies in ultrasound elastography of (a) the liver and, (b) the kidney, (c) the estimated ratio of the SWS of the kidney to the SWS of the liver.

### 4.3. Magnetic resonance elastography

To assess the capability of AIA in visualizing the lesions in MRE scans, our proposed estimator was applied to the 3D MRE datasets obtained from the brain-mimicking phantom experiment. By leveraging the autocorrelation approaches, our objective was to extract the SWS map and use it for lesion visualization and characterization. This study specifically focused on the X-motion and Z-motion signals, which exhibited the strongest signal strengths in this configuration. Utilizing these MRE experimental data, SWS maps were estimated by employing simple autocorrelation approaches in the $x$ and $y$ directions, as well as the AIA approach. Figure 11 presents the estimated SWS within the brain-mimicking phantom at an excitation frequency of 200 Hz. Figures 11(a) and (b) display the estimated SWS for the X-motion MRE data using the simple baseline autocorrelation in the $x$-direction and $y$-direction respectively. Figure 11(c) showcases the estimated SWS using AIA for X-motion MRE data. Similarly, figures 11(d), (e), and (f) represent the estimated SWS for the Z-motion MRE data using the simple baseline autocorrelation in the $x$ and $y$ directions, and the AIA approach, respectively. It is worth noting that a square autocorrelation window with dimensions of 10.7 mm was used in these measurements.

The results demonstrate that the simple autocorrelation in the $x$-direction fails to differentiate between lesions within the brain phantom, both for the X-motion displacement field (see figure 11(a)), and the Z-motion displacement field (see figure 11(d)). The application of simple autocorrelation in the $y$-direction on the X-motion dataset (see figure 11(b)) leads to the visualization of two lesions in the background material. However, in the Z-motion dataset (see figure 11(e)) only one of the lesions is visualized along with two artificial unrealistic lesions. In contrast, as evident in both figures 11(c) and (f), the estimated SWS maps using AIA for the X-motion and Z-motion displacement fields successfully visualize the presence of two lesions within the brain-mimicking phantom. Furthermore, the left-side lesion (18 mm diameter) appears as a larger zone compared to the right-side lesion (12 mm diameter), indicating that the left-side lesion is larger in size than the right-side lesion.

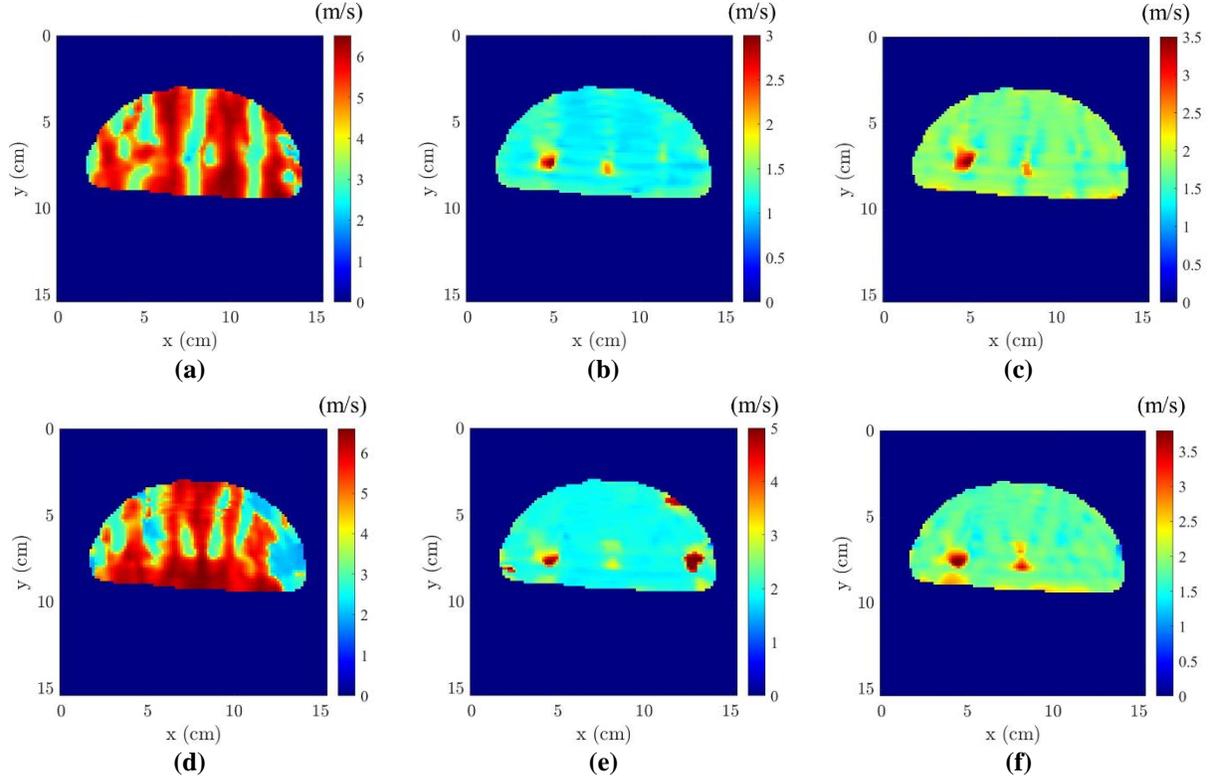

**Figure 11.** Estimated SWS of the brain-mimicking phantom with two lesions at the frequency of 200 Hz using (a) simple autocorrelation in *x*-direction, (b) simple autocorrelation in *y*-direction, and (c) the AIA approach on the X-motion MRE dataset. Estimated SWS using (d) simple autocorrelation in *x*-direction (e) simple autocorrelation in *y*-direction, and (f) the AIA approach on the Z-motion MRE dataset.

Table 7 presents the SWS for the MRE brain phantom in the background material, estimated using simple autocorrelation approaches and the AIA approach in X-motion and Z-motion displacement fields. All MRE measurements of SWS and SNR were conducted considering the almost full 2D background with some distance to the lesions' boundaries. According to Kabir *et al* (2023), the background shear modulus for the brain phantom is 3.34 ± 0.04 kPa, equivalent to SWSs of 1.83 m/s. It is evident that the estimated SWS using $k_x$ is an outlier with an average error of 157% for the X-motion and Z-motion displacement fields in the background. This error is 25%, using $k_y$. In contrast, the average error for estimating SWS using AIA in the background is 2%.

**Table 7.** SWS in the background material for the brain phantom MRE estimated using simple autocorrelation approaches and the AIA approach in X-motion and Z-motion displacement fields, errors were calculated using the ground truth SWS value of 1.83 (m/s) for the brain phantom as reported in Kabir et al (2023).

| Displacement field | SWS (m/s) in background using simple autocorrelation ($k_x$) | SWS (m/s) in background using simple autocorrelation ($k_y$) | SWS (m/s) in background using AIA | SWS estimation error using simple autocorrelation ($k_x$) | SWS estimation error using simple autocorrelation ($k_y$) | SWS estimation error using AIA |
|---|---|---|---|---|---|---|
| X-motion | 4.80 ± 1.22 | 1.19 ± 0.13 | 1.76 ± 0.15 | 162% | 35% | 4% |
| Z-motion | 4.59 ± 1.44 | 2.09 ± 0.48 | 1.83 ± 0.22 | 151% | 14% | 0% |

Table 8 illustrates the SNR of estimated SWS within the background material for the brain phantom MRE using simple autocorrelation approaches in the *x* and *y* directions, and the AIA approach for both X-motion and Z-motion displacement fields. These results clearly demonstrate a substantial enhancement in using the AIA approach compared to simple autocorrelation approaches.

**Table 8.** SNR of estimated SWS in the background material for the brain phantom MRE using simple autocorrelation approaches and the AIA approach in X-motion and Z-motion displacement fields.

| Displacement field | SNR (dB) in background using simple autocorrelation ($k_x$) | SNR (dB) in background using simple autocorrelation ($k_y$) | SNR (dB) in background using AIA | AIA improvement over $k_x$ | AIA improvement over $k_y$ |
|---|---|---|---|---|---|
| X-motion | 5.96 | 9.71 | 10.66 | 79% | 10% |
| Z-motion | 5.03 | 6.43 | 9.15 | 82% | 42% |

## 4.4. Optical coherence elastography

In the OCE experiment, a 3D data set was obtained depicting the displacement field in the optical direction of the system, denoted as the *z*-direction in this study. The SWS maps for the phantom including a stiff lesion were determined by employing simple autocorrelation estimators in the *x* and *y* directions, along with the AIA approach. In this case study, the autocorrelation window size was adjusted to 1.3 mm × 1.3 mm.

Figure 12 presents the SWS obtained from the simple *x*-direction autocorrelation (see figure 12(a)), simple *y*-direction autocorrelation (see figure 12(b)), and the AIA approach (see figure 12(c)) for the phantom featuring a stiff lesion at the frequency of 1500 Hz. The lesion's location and size are outlined by black dashed lines in the center of the phantom. As shown in figure 12, the AIA approach effectively visualizes the stiff lesion as a distinct red region within the soft background. In contrast, the simple autocorrelation approaches encountered challenges in accurately identifying the stiff lesion. Additionally, simple autocorrelation approaches resulted in a significant level of uncertainty in SWS estimations within the background region.

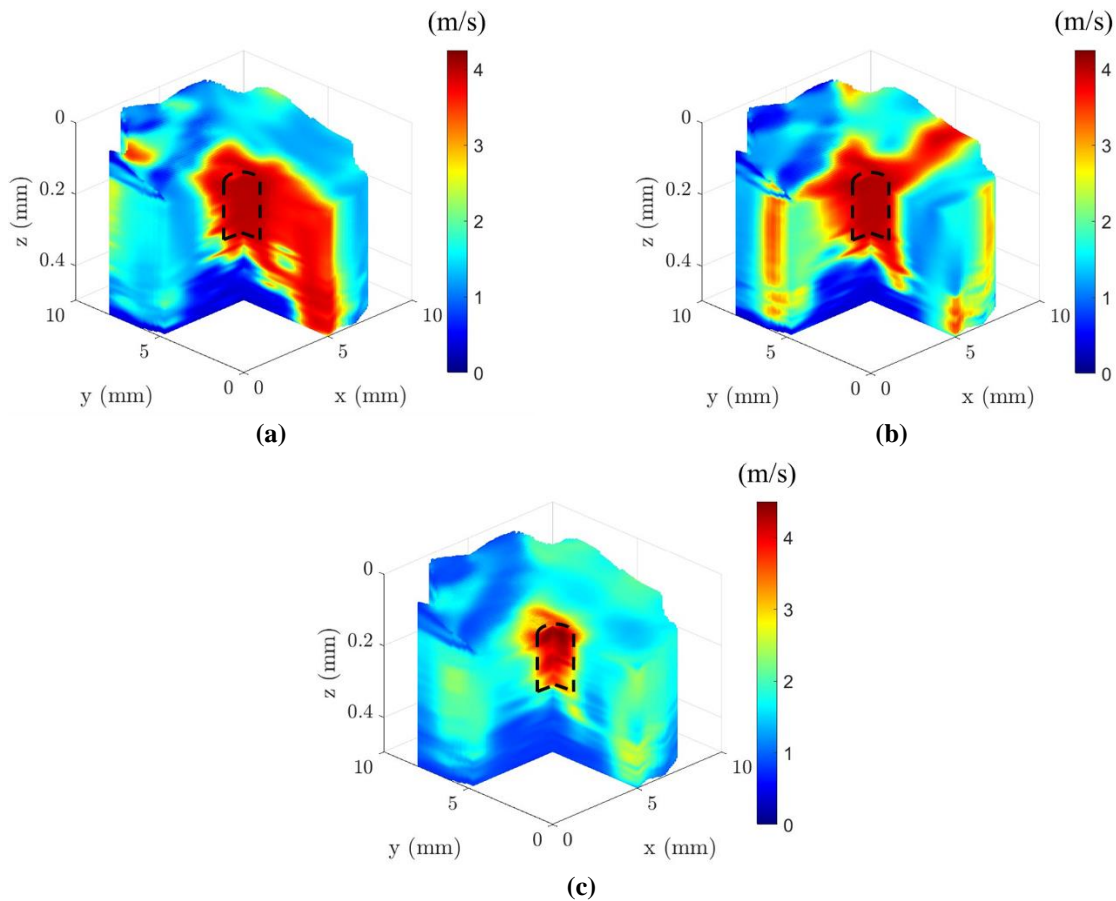

**Figure 12.** Comparative evaluation of estimated SWS for the OCE scan of the gelatin phantom with a lesion at the frequency of 1500 Hz using (a) simple autocorrelation approach in *x*-direction, (b) simple autocorrelation approach in *y*-direction, and (c) the AIA approach.

Table 9 displays the comparison of SWS within the background material of the gelatin phantom estimated using simple autocorrelation techniques and the AIA approach. The estimated SWS and SNR in OCE analysis were computed considering the almost full 3D background with some distance to the boundaries. Errors were computed using the established ground truth value of 1.94 (m/s) for the SWS of the 5% gelatin phantom, as documented in Zvietcovich et al (2019a).

**Table 9.** SWS in the background material the gelatin phantom OCE estimated using simple autocorrelation approaches and the AIA approach, errors were calculated using the ground truth SWS value of 1.94 (m/s) for a 5% gelatin phantom as reported in Zvietcovich et al (2019a).

| Excitation frequency | SWS (m/s) in background using simple autocorrelation ($kx$) | SWS (m/s) in background using simple autocorrelation ($ky$) | SWS (m/s) in background using AIA | SWS estimation error using simple autocorrelation ($kx$) | SWS estimation error using simple autocorrelation ($ky$) | SWS estimation error using AIA |
|---|---|---|---|---|---|---|
| 1500 Hz | 2.43 | 2.28 | 2.05 | 25% | 18% | 6% |

Table 10 displays the SNR of estimated SWS in the background material for the gelatin phantom OCE at the frequency of 1500 Hz utilizing simple autocorrelation approaches and the AIA approach. This table highlights the substantial improvement offered by AIA over simple autocorrelation approaches with a 117% improvement over *kx* and a 127% improvement over *ky*. These results highlight the exceptional capabilities of the AIA approach in effectively identifying and characterizing stiff lesions within the gelatin phantom in OCE experiments.

**Table 10.** SNR of estimated SWS in the background material for the gelatin phantom OCE using simple autocorrelation approaches and the AIA approach.

| Excitation frequency | SNR (dB) in background using simple autocorrelation (*kx*) | SNR (dB) in background using simple autocorrelation (*ky*) | SNR (dB) in background using AIA | AIA improvement over *kx* | AIA improvement over *ky* |
|---|---|---|---|---|---|
| **1500 Hz** | **4.91** | **4.69** | **10.66** | **117%** | **127%** |

## 5. Conclusion

This study introduced the AIA estimator, developed explicitly for elastography measurements in reverberant shear wave fields. Through the integration of data from all directions of the 2D autocorrelation function, the AIA method exhibited robustness in the presence of data noise. We note that the 1D AIA approach is more computationally efficient than alternative 2D curve fitting to the theoretical autocorrelation function in any imaging plane, where 2D computations increase as $N^2$ points within the window as opposed to N points for AIA. Furthermore, the $N^2$ fit has a greater potential for local minima within the mean squared error profile. Finally, as shown in figure 8(b), the autocorrelation function in imperfect reverberant fields (having a dominant direction) can have an anomalous pattern in one direction that does not readily converge to the theoretical curve, and this is ameliorated in the AIA approach.

A comprehensive examination was conducted to assess the effectiveness of this advanced autocorrelation approach, using a numerical simulation featuring a stiff branching tube in a uniform background. The AIA approach demonstrated enhanced accuracy in estimating the SWS ratio between the stiff branching tube and the background material. Furthermore, the SNR of the estimated SWS within the background material illustrates the improvement achieved with the AIA approach compared to simple autocorrelation approaches. The practical performance of the AIA approach was assessed through a series of experiments, including ultrasound elastography of a breast phantom with a lesion, ultrasound elastography of the liver-kidney region, MRE of a brain-mimicking phantom with two lesions, and OCE of a gelatin phantom with a stiff lesion. Across this diverse array of experiments, shear wave fields, and excitation frequencies, the AIA results consistently demonstrated substantial enhancements in the estimated SWS as well as the SNR values. This study has presented the AIA estimator as a robust tool for improving SWS estimation in SWE, offering enhanced accuracy across a spectrum of applications, even in the presence of imperfect reverberant shear wave fields. These findings highlighted the potential of the AIA approach to advance the field of elastography and contribute to more precise characterization of the elastic properties of tissue in clinical applications.

## Acknowledgments

This work was supported by the University of Rochester Center of Excellence in Data Science for Empire State Development. We are grateful to Elastance Imaging, Inc. and Dr. Richard Barr for sharing a scan from their previous studies.

## Appendix

In this Appendix, we provide additional details, comparisons, and metrics for the AIA approach. First, the simulation is examined with focus on the accuracy of the AIA estimator compared to the average of simple autocorrelation approaches (tables A.1 and A.2). Other comparisons to study the impact of added noise on the SWS estimation are summarized in table A.3.

**k-Wave simulation**

Figure A.1 displays the average SWS estimated using single autocorrelation approaches in different directions (left column in Figure 7) in k-Wave simulation at the frequency of 200 Hz. Table A.1 presents estimated SWS in both the background material and the branching tube along with the SWS ratio using an average of simple autocorrelation approaches in different autocorrelation planes. As indicated in this table, the AIA approach provides an estimation of the SWS ratio, with an average error of just 4%. In contrast, the estimated SWS ratio using an average of simple autocorrelation approaches, demonstrates an average error of over 14%.

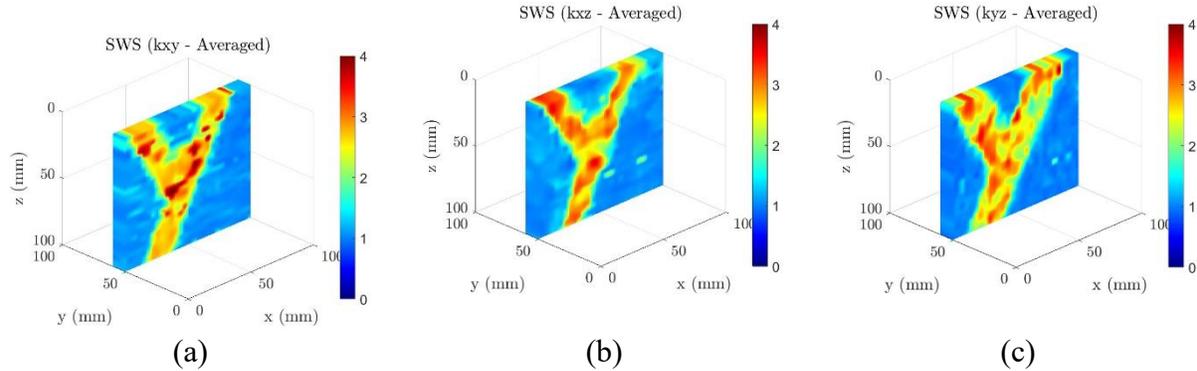

(a)          (b)          (c)

**Figure A.1.** Average SWS estimated using the simple autocorrelation approaches in k-Wave simulation at the frequency of 200 Hz (a) average of $k_x$ and $k_y$ in the $xy$ plane, (b) average of $k_x$ and $k_z$ in the $xz$ plane, and (c) average of $k_y$ and $k_z$ in the $yz$ plane.

**Table A.1.** Comparison of SWS in the background material and the branching tube, and the SWS ratio estimated using the average of simple autocorrelation approaches across different autocorrelation planes.

| Autocorrelation plane | SWS (m/s) in background using an average of simple autocorrelation approaches | SWS (m/s) in branching tube using an average of simple autocorrelation approaches | SWS ratio using an average of simple autocorrelation approaches | SWS ratio estimation error using an average of simple autocorrelation approaches | SWS ratio estimation error using AIA |
|---|---|---|---|---|---|
| $xy$ plane | 1.08 ± 0.08 | 2.80 ± 0.41 | 2.59 ± 0.43 | 14% | 3% |
| $xz$ plane | 1.08 ± 0.12 | 2.72 ± 0.41 | 2.52 ± 0.47 | 16% | 7% |
| $yz$ plane | 1.02 ± 0.08 | 2.66 ± 0.35 | 2.61 ± 0.40 | 13% | 2% |

Table A.2 depicts a comparison of the SNR of estimated SWS in the background material using the average of simple autocorrelation approaches and the AIA approach across different autocorrelation planes.

**Table A.2.** SNR comparison of estimated SWS in the background material using the average of simple autocorrelation approaches and the AIA approach across different autocorrelation planes.

| Autocorrelation plane | SNR (dB) in background using the average of simple autocorrelation approaches | SNR (dB) in background using AIA | AIA improvement over simple autocorrelation |
|---|---|---|---|
| *xy* plane | 11.40 | 13.69 | 20% |
| *xz* plane | 9.53 | 9.68 | 2% |
| *yz* plane | 10.80 | 10.98 | 2% |

In order to assess the impact of added noise on the SWS estimation, different levels of white Gaussian noise including 10 dB SNR, 1 dB SNR, -1 dB SNR, and -10 dB SNR were introduced into the k-Wave simulation of the stiff branching tube in the uniform background. Subsequently, three-dimensional median filtering was applied to the velocity field. Figure A.2 illustrates the velocity field both before filtering and after filtering, showcasing the impact of different noise levels.

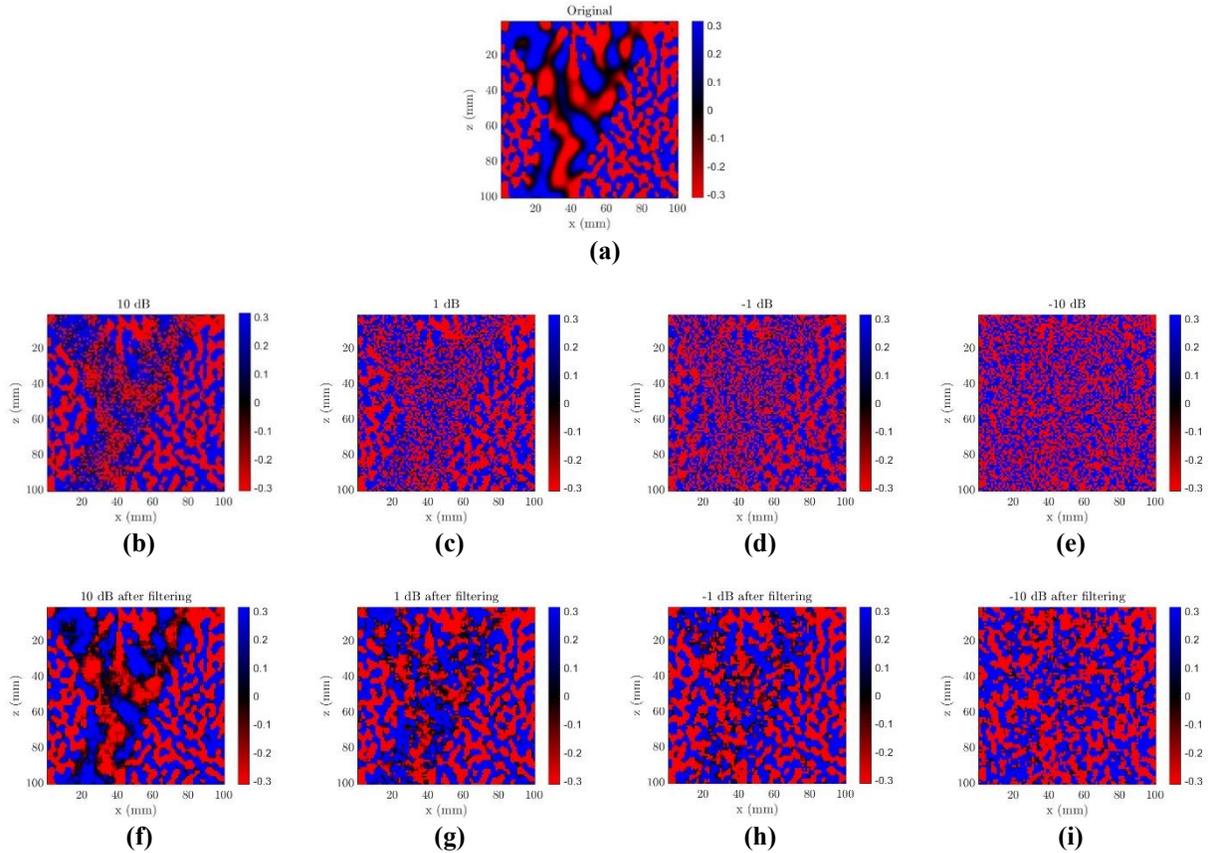

**Figure A.2.** Velocity field of the stiff branching tube in the uniform background at the frequency of 200 Hz across different levels of added white noise, (a) without noise, (b) with a 10 dB SNR, (c) with 1 dB SNR, (d) with -1 dB SNR, (e) with -10 dB SNR, (f) with 10 dB SNR after filtering, (g) with 1 dB SNR after filtering, (h) with -1 dB SNR after filtering, (i) with -10 dB SNR after filtering.

Figure A.3. presents the SWS estimated using the AIA approach at the frequency of 200 Hz across different autocorrelation planes and different levels of added white noise. It is worth noting that the estimated SWS for the noise level of 10 dB is presented in Figure 7. The SWS estimated in both the background material and the branching tube using the AIA approach are compared across different levels of added Gaussian white noise and different autocorrelation planes in table A.3. As shown, the estimated SWS using AIA remains reasonable even in the presence of a high level of noise, e.g., -10 dB SNR.

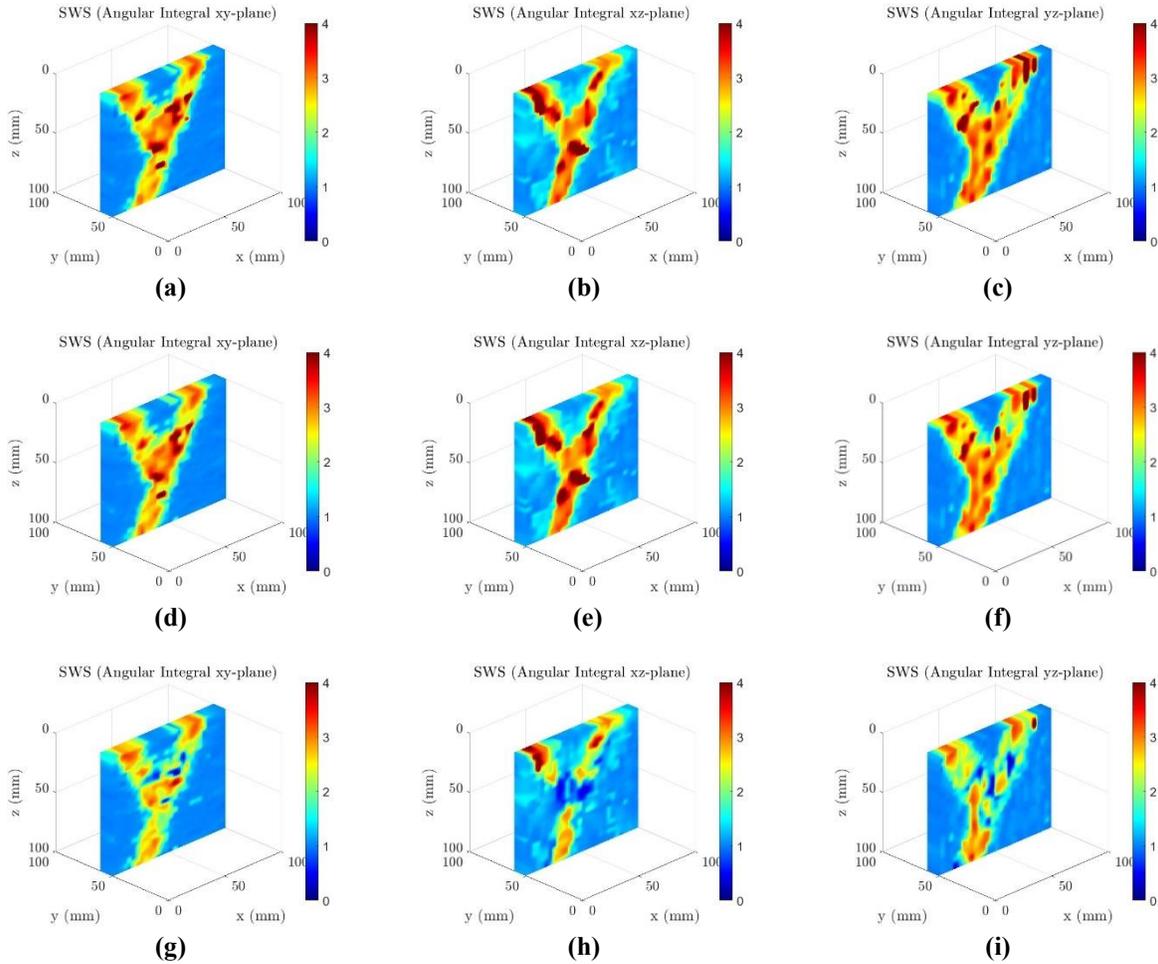

**Figure A.3.** SWS estimated using the AIA approach at the frequency of 200 Hz across different autocorrelation planes and different levels of added noise (a) using AIA in xy-plane with 1 dB added noise, (b) using AIA in xz-plane with 1 dB added noise, (c) using AIA in yz-plane with 1 dB added noise, (d) using AIA in xy-plane with -1 dB added noise, (e) using AIA in xz-plane with -1 dB added noise, (f) using AIA in yz-plane with -1 dB added noise, (g) using AIA in xy-plane with -10 dB added noise, (h) using AIA in xz-plane with -10 dB added noise, (i) using AIA in yz-plane with -10 dB added noise.

**Table A.3.** Comparison of SWS both in the background material and the branching tube estimated using the AIA approach in different levels of added Gaussian white noise and different autocorrelation planes.

| Added Gaussian white noise | Autocorrelation plane | SWS (m/s) in background using AIA | SWS (m/s) in branching tube using AIA | SWS ratio using AIA | SWS ratio estimation error using AIA | SWS ratio estimation error using an average of simple autocorrelation approaches |
|---|---|---|---|---|---|---|
| 1 dB | xy plane | 1.03 ± 0.05 | 2.92 ± 0.55 | 2.83 ± 0.55 | 6% | 17% |
|  | xz plane | 1.15 ± 0.12 | 3.01 ± 0.72 | 2.62 ± 0.68 | 13% | 21% |
|  | yz plane | 1.06 ± 0.08 | 2.95 ± 0.57 | 2.78 ± 0.58 | 7% | 17% |
| -1 dB | xy plane | 1.03 ± 0.06 | 2.89 ± 0.50 | 2.81 ± 0.51 | 6% | 18% |
|  | xz plane | 1.15 ± 0.12 | 3.09 ± 0.73 | 2.69 ± 0.69 | 10% | 22% |
|  | yz plane | 1.06 ± 0.08 | 2.90 ± 0.53 | 2.74 ± 0.54 | 9% | 17% |
| -10 dB | xy plane | 1.03 ± 0.07 | 2.38 ± 0.59 | 2.31 ± 0.59 | 23% | 37% |
|  | xz plane | 1.13 ± 0.11 | 1.98 ± 0.88 | 1.75 ± 0.80 | 42% | 52% |
|  | yz plane | 1.05 ± 0.07 | 2.16 ± 0.71 | 2.06 ± 0.69 | 31% | 41% |